\documentclass[aps,pre,twocolumn,showpacs]{revtex4}
\usepackage{graphicx}

\begin{document}

\title{Network algorithmics and the emergence of information integration in
cortical models}

\author{Andre Nathan}
\author{Valmir C. Barbosa}
\affiliation{Programa de Engenharia de Sistemas e Computa\c c\~ao, COPPE,
Universidade Federal do Rio de Janeiro,
Caixa Postal 68511, 21941-972 Rio de Janeiro - RJ, Brazil}

\begin{abstract}
An information-theoretic framework known as integrated information theory (IIT)
has been introduced recently for the study of the emergence of consciousness in
the brain [D.~Balduzzi and G.~Tononi, PLoS Comput.\ Biol.\ \textbf{4}, e1000091
(2008)]. IIT purports that this phenomenon is to be equated with the generation
of information by the brain surpassing the information which the brain's
constituents already generate independently of one another. IIT is not fully
plausible in its modeling assumptions, nor is it testable due to severe
combinatorial growth embedded in its key definitions. Here we introduce an
alternative to IIT which, while inspired in similar information-theoretic
principles, seeks to address some of IIT's shortcomings to some extent. Our
alternative framework uses the same network-algorithmic cortical model we
introduced earlier [A.~Nathan and V.~C.~Barbosa, Phys.\ Rev.~E \textbf{81},
021916 (2010)] and, to allow for somewhat improved testability relative to IIT,
adopts the well-known notions of information gain and total correlation applied
to a set of variables representing the reachability of neurons by messages in
the model's dynamics. We argue that these two quantities relate to each other in
such a way that can be used to quantify the system's efficiency in generating
information beyond that which does not depend on integration, and give
computational results on our cortical model and on variants thereof that are
either structurally random in the sense of an Erd\H{o}s-R\'{e}nyi random
directed graph or structurally deterministic. We have found that our cortical
model stands out with respect to the others in the sense that many of its
instances are capable of integrating information more efficiently than most of
those others' instances.
\end{abstract}


\maketitle

\section{Introduction}
\label{sec:intro}

Explaining the emergence of consciousness out of the massive neuronal
interactions that take place in the brain is the greatest unsolved problem in
neuroscience. It defies our ability to define what it means for the brain to be
in a conscious state and, lacking a definition, also our ability to pinpoint the
mechanisms that give rise to such a state and its evolution. The most recent
player in the quest for a framework for consciousness studies is the integrated
information theory (IIT) \cite{balduzzi08}. This theory is information-theoretic
in nature and seeks to characterize consciousness on the formal grounds of how
information originating at different parts of the brain gets integrated as
neurons interact with one another. IIT has been met with enthusiasm (cf., e.g.,
\cite{k09}), but substantial further developments are needed to help clarify
whether this is justified.

IIT is defined on a directed graph having a node for each of a group of
variables. These variables' values evolve in lockstep (i.e., in discrete time,
much as in a cellular automaton \cite{i01}) in such a way that, at time $t+1$,
the value of a particular variable is a function of its own value and of those
of its in-neighbors in the graph at time $t$. Each node is thus assumed to have
a local function that it applies on inputs to get an output per time unit.
Typically a variable's possible values are $0$ or $1$ and a node's local
function is one of the elementary logical operations. The basic tenet of IIT is
that consciousness is to be equated with the surplus of information that the
system is capable of generating, relative to the total information that is
generated by its parts independently of one another, as it evolves from an
initial state of maximum uncertainty to a final state. The use of ``parts'' here
refers to a specific partition of the set of variables. The information surplus
that IIT considers is the minimum over all possible partitions.

Unless a lot of regularity is present, computing this minimum information
surplus requires a number of partitions to be examined that is given in the
worst case by the Bell number corresponding to the number of variables. The Bell
number for as few as $20$ variables, say, is already of the order of $10^{13}$
\cite{oeis}, so the task at hand is computationally intractable even for
modestly sized systems. Another potential obstacle to the success of IIT in
eventually fulfilling the promise of helping characterize the emergence of
consciousness is the apparent oversimplistic character of some of its elements.
In our view, these include the use of binary variables and operations (even if
probabilistic), and also the assumption that the system evolves in time in a
synchronous and memoryless fashion.

Here we study the emergence of information integration while striving both to
adhere to the spirit of IIT and to address its potential shortcomings. The main
elements of our approach are the following.

(a) We adopt a cortical model with ample provisions for randomness, asynchrony,
and neuronal memory. This model is the same we used previously in \cite{nb10},
having a structural component and a functional one. The structural component is
a random directed graph and attempts to portray, to the fullest possible extent,
whatever structural characteristics cortices can at present be said to have. The
model's functional component, in turn, prescribes a randomized distributed
algorithm to run on the graph. This algorithm uses message passing to mimic
inter-neuron signaling over synapses. It is also fully asynchronous, meaning
that local actions are triggered by message arrivals independently of what is
happening elsewhere in the graph. At each node, the algorithm is capable of
providing enough bookkeeping for some of the node's history to be influential on
current actions.

(b) We adopt two interrelated indicators of information integration. The first
is simply information gain, that is, the amount of information the system
generates as a single entity from an initial situation of maximum uncertainty.
The second is total correlation, which in essence indicates by how much the
first indicator surpasses the nodes' total gain of information when each node's
gain is considered independently of all others'. Unlike IIT, the variables
involved in our approach do not bear directly on the fire/hold dichotomy of
binary local states but on whether the corresponding nodes are reached by
messages as the computation unfolds.

The distributed algorithm in (a) requires at least one node to behave
non-reactively at the beginning of a run. That is, at least one node must have
the chance to send messages out spontaneously without any incoming message to
trigger its actions. The algorithm admits any number of such initiators to act
concurrently. Choosing the nodes to do it in each run is a random process.
Together with the randomness already present in the graph's structure and in the
algorithm, this random choice of initiators leads to an assessment of the
indicators described in (b) that takes place by averaging them over a number of
graphs and/or a number of runs on each graph, each run with a new set of
initiators. Vis-\`{a}-vis the treatment of information integration in IIT, which
makes reference to an optimal partition of the set of variables, our approach is
to generate a great number of message-flow patterns and to measure information
integration as an expected, rather than optimal, quantity.

We regard the present work as being fully in line with several others that have
recently attempted to draw on graph-based methods to help solve problems in
neuroscience
\cite{sporns04,sporns05,achard06,bassett06,he07,honey07,reijneveld07,sporns07,stam07,yhsn08}.
These works are all based on highly abstract models of the underlying biological
system, but some researchers believe that a complete understanding of the
system's properties can only come from considering every possible detail, even
down to the molecular level. So a sort of methodological chasm is beginning to
appear, as documented in the news item found in \cite{l10}. As in the present
work's predecessor \cite{nb10}, here we adopt what might be called the
artificial-life stance \cite{f00}, which essentially posits the middle
alternative of employing only as much modeling detail as required to let some
``life as it could be'' properties emerge. Vague though this sounds, the
cortical model we use has been shown to give rise to some such properties
\cite{nb10}. Specifically, by relying on the combination of its two main
components (one structural, the other functional), our model gives rise, with
excellent agreement, to experimentally obtained lognormal distributions of
synaptic strengths \cite{song05}. Moreover, by including enough detail of
inter-neuron signaling so that the all-important local histories
\cite{barbour07} can always be retrieved for careful examination, our model also
reveals signs of the very rich dynamics that everyone agrees must underlie all
cortical functions.

We proceed according to the following layout. The two components of our cortical
model are reviewed in Secs.~\ref{sec:model} and~\ref{sec:algo}. Then we move, in
Sec.~\ref{sec:infoint}, to a description of the information-integration
indicators to be used. We give computational results in Sec.~\ref{sec:results}
and follow these with discussion and conclusions, in Secs.~\ref{sec:disc}
and~\ref{sec:concl}, respectively.

\section{Network model}
\label{sec:model}

The structural portion of our cortical model is the same as in \cite{nb10}. It
consists of a directed graph $D$ having $n$ nodes, one for each neuron. In $D$,
an edge leading from node $i$ to node $j$ indicates that a synapse exists
between the axon of the neuron that node $i$ represents and one of the dendrites
of the neuron represented by node $j$. The existence of such an edge, therefore,
amounts to the possibility of direct causal influence of what happens at node
$i$ upon what happens at node $j$. In the same vein, indirect causal influence
of what happens at node $i$ upon what happens at farther nodes can also exist,
provided those nodes can be reached from $i$ through directed paths of $D$. If
$i$ is part of any directed cycle in $D$, then it follows that present events at
node $i$ can causally influence future events at the same node also through the
indirect mediation of all other nodes in the cycle.

We regard $D$ as a originating from a random-graph model, so completing its
definition requires that we specify how the out-degree of a randomly chosen node
(its number of out-neighbors) is distributed, and also the probability that one
of these out-neighbors is another randomly chosen node (this, indirectly,
specifies the distribution of a randomly chosen node's in-degree, its number of
in-neighbors). Still following \cite{nb10}, we assume that a randomly chosen
node has out-degree $k>0$ with probability proportional to $k^{-1.8}$. The
adoption of a scale-free law \cite{newman05} with this particular exponent
follows the work in \cite{eguiluz05,vandenheuvel08}, but one should mind the
caveat given below. If $i$ is the randomly chosen node in question, what is left
to specify is the probability that each out-neighbor of $i$ is precisely another
randomly chosen node, say $j$. Taking inspiration from the work in
\cite{kaiser04a,kaiser04b}, first we assume that the nodes of $D$ are placed
uniformly at random on a radius-$1$ sphere. If $d_{ij}$ is the resulting
Euclidean distance between $i$ and $j$, then each out-neighbor of $i$ coincides
with $j$ with probability proportional to $e^{\lambda d_{ij}}$, with $\lambda<0$
a constant. This constant affects the size of $D$'s giant strongly connected
component (GSCC) \cite{dorogovtsev01} heavily. The GSCC of $D$ is the largest
subgraph of $D$ in which a directed path exists between any two nodes, that is,
the largest subgraph in which all nodes have the potential of exerting direct or
indirect causal influence upon all others. Similarly to what is explained in
\cite{nb10}, we choose $\lambda=-1$ so that the expected number of nodes in the
GSCC is about $0.9n$. Henceforth, we limit all our information-integration
investigations to within the GSCC of $D$.

The aforementioned caveat is the following. Even though the graph-theoretic
modeling of cortices has made great progress \cite{bs09} since the earliest
attempts (as represented, e.g., by \cite{abeles91}, where the random graphs of
Erd\H{o}s and R\'{e}nyi \cite{erdos59} were used directly), our adoption of a
scale-free structure is far from any form of consensus. This is not to say that
cortices have no scale-free properties: in fact, it has been argued that they do
indeed have such properties \cite{yhsn08,fkbr09,s11}, including the so-called
small-world characteristics \cite{asbs00}, the presence of hubs (nodes with a
great number of out-neighbors), and many others \cite{f07}. What is meant,
instead, is that there exist results pointing in contradictory directions. The
recent growth model in \cite{fkbr09}, for example, gives rise to an out-degree
distribution that is not scale-free. If, on the other hand, we concede that this
model is not fully justified biologically and look instead for topological
characteristics derived from measurements on real cortices, what we find is not
at the level of detail that we need (i.e., the level of neuronal wiring). Our
sources for the $k^{-1.8}$ power law \cite{eguiluz05,vandenheuvel08}, for
example, adopt the granularity of functional parts of the cortex. The latest
available mapping \cite{ms10} (in fact the most comprehensive to date), in
contrast, adopts structural (rather than functional) granularity and leads to
the conclusion of an exponential (rather than a power-law) distribution.

\begin{figure}
\centering
\scalebox{0.50}{\includegraphics{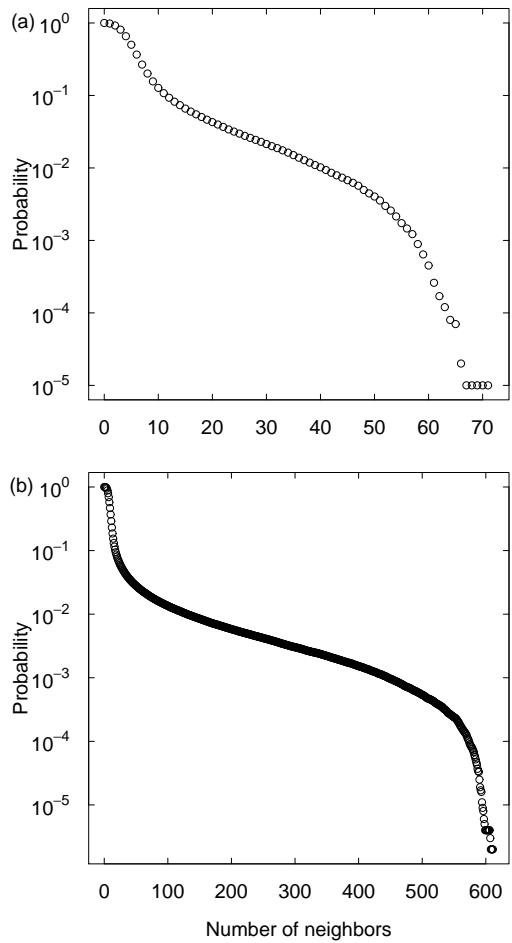}}
\caption{Distribution of a node's number of neighbors (in- and out-neighbors
combined) in our cortical model for $n=100$ (a) and $n=1\,000$ (b).
Probabilities are shown for the complementary cumulative distribution; that is,
given a number $k>0$ of neighbors, we show the probability that a randomly
chosen node has strictly more than $k$ neighbors. Data are averages over
$1\,000$ graphs for each value of $n$.}
\label{fig:figure1}
\end{figure}

Another problem with these measurement-based characterizations is that all the
reported distributions are in fact distributions of degrees, not out-degrees.
That is, what counts for each node is its total number of neighbors (in- and
out-neighbors combined). While in \cite{nb10} we ignored this and adopted the
power law of \cite{eguiluz05,vandenheuvel08} as the only measurement-based
distribution available at the time, the more recent results in \cite{ms10} lend,
somewhat surprisingly, new support to our choice of this power law. In fact, we
have found empirically that the degree distribution that results from our
assumed out-degree distribution and distance-based deployment of directed edges
can be approximated by an exponential over a significant range of degrees. This
can be seen in Fig.~\ref{fig:figure1}, where the distribution of the combined
in- and out-degrees of the nodes of $D$ is shown. In our view, this provides all
the justification we can have at this point regarding our choice of the
random-graph model. Further justification (or, more likely, adaptation) will
depend on wiring-level measurements, whose availability, to the best of our
knowledge, is still not foreseeable.

\section{Network algorithmics}
\label{sec:algo}

In order to fully describe the cortical model we use in the present study, the
structural properties of graph $D$ given in Sec.~\ref{sec:model} need to be
complemented with further, functional properties of the graph's nodes and edges.
A goal of the resulting model is to provide an algorithmic abstraction of
cortical functioning that can mimic, to some extent, the buildup of potential at
each neuron as its dendrites are reached by action potentials traveling down
other neurons' axons, as well as the eventual firing that this buildup entails
with the accompanying action potential that travels down the neuron's own axon.
Another goal is to simulate the dynamics of synaptic strengths as they vary in
the wake of neuronal firing.

We provide the necessary functional component of the model in the form of an
asynchronous distributed algorithm \cite{barbosa96}. In general, such an
algorithm assumes that the nodes of $D$ are capable of receiving messages from
their in-neighbors in the graph, of performing local computation on the
information that is thus received, and finally of sending messages to their
out-neighbors. This processing may occur concomitantly at several of the graph's
nodes and edges and this is what gives the algorithm its distributed character.
What gives it its asynchronous character, in turn, is the underlying assumption
that the local computation at each node makes no reference to temporal
quantities of a global nature. Such inherent locality gives asynchronous
distributed algorithms a clear connection to most complex-network studies of the
past decade, as evinced by the various contributions collected in
\cite{bs03,nbw06,bkm09}. Surprisingly, though, to the best of our knowledge the
interplay of structure and function, and its role in giving rise to global
network properties, has seldom been explored. One notable example is the work
in \cite{sb06}, where efficient information dissemination is shown to emerge
from strictly local decisions. Another example is provided by the present study
and its predecessor \cite{nb10}.

The algorithm we use is the one introduced in \cite{nb10}. As mentioned in
Sec.~\ref{sec:intro}, here we assume that algorithm to be sufficiently certified
for use in the cortical model we study because it is in \cite{nb10} shown to be
capable of giving rise, among other things, to experimentally observed
distributions of synaptic strengths. This algorithm, henceforth referred to
simply as $A$, uses $v_j$ to represent the potential of node $j$ and $w_{ij}$ to
represent the synaptic strength of the edge directed from node $i$ to node $j$.
As customary, we refer to each $w_{ij}$ as a synaptic weight. We also assume
that all nodes share the same rest and threshold potentials, denoted by $v^0$
and $v^\mathrm{t}$ (with $v^0<v^\mathrm{t}$), respectively, and that every
synaptic weight lies in the interval $[0,1]$. Finally, nodes can be excitatory
or inhibitory and $D$ contains no edge connecting two inhibitory nodes
\cite{abeles91}.

Our specification of algorithm $A$ is based on the following procedure, called
$\textit{Fire}$, which gives a probabilistic rule for node $j$ to fire, that is,
to send messages to its out-neighbors and reset its potential to the rest
potential. Each of these messages is to be interpreted as the signaling by $j$
to one of its out-neighbors, through the corresponding synapse, that results
from node $j$'s firing and the ensuing action potential. Procedure
$\textit{Fire}$ uses a probability parameter, $p$.

\bigskip\noindent
\textsc{Procedure $\textit{Fire}$:}

\noindent
Fire with probability $p$:
\begin{enumerate}
\item Send a message to each out-neighbor of node $j$.
\item Set $v_j$ to $v^0$.
\end{enumerate}

Node $j$ participates in a run of algorithm $A$ through a series of calls to
procedure $\textit{Fire}$ with suitable $p$ values. In the first call in this
series node $j$ may be one of the so-called initiators of the run. In this case
its participation is restricted to calling $\textit{Fire}$ with $p=1$. All calls
in which $j$ is not an initiator are reactive, in the sense that node $j$ only
computes upon receiving a message from some in-neighbor. In this case, the call
to $\textit{Fire}$ is part of a larger set of rules, as given next when node $i$
is the in-neighbor in question. In this larger set, the call to procedure
$\textit{Fire}$ is preceded by an alteration to the potential $v_j$ and
followed, possibly, by an alteration to the synaptic weight $w_{ij}$.

\bigskip\noindent
\textsc{Algorithm $A$ (reactive mode):}

\begin{enumerate}
\item If $i$ is excitatory, then set $v_j$ to $\min\{v^\mathrm{t},v_j+w_{ij}\}$.
\item If $i$ is inhibitory, then set $v_j$ to $\max\{v^0,v_j-w_{ij}\}$.
\item Call procedure $\textit{Fire}$ with $p=(v_j-v^0)/(v^\mathrm{t}-v^0)$.
\item If firing did occur during the execution of $\textit{Fire}$, then set $w_{ij}$ to
$\min\{1,w_{ij}+\delta\}$.
\item If firing did not occur during the execution of $\textit{Fire}$ but the previous
message received by node $j$ from any of its in-neighbors did cause $j$ to fire,
then set $w_{ij}$ to $(1-\alpha)w_{ij}$.
\end{enumerate}

In steps 4 and 5 of algorithm $A$, $\delta>0$ and $\alpha$ such that
$0<\alpha<1$ are parameters meant to let the algorithm follow, though only to a
limited extent, the principles of spike-timing-dependent plasticity
\cite{song00,abbott00}. These principles dictate, as a general rule, that the
synaptic weight is to increase if firing occurs, decrease otherwise, always as
a function of how close in time the relevant firings by nodes $i$ and $j$ are.
Moreover, increases are to occur by a fixed amount, decreases by proportion
\cite{bi98,bi01,kepecs02}. As explained in \cite{nb10}, $\delta$ and $\alpha$
must be such that $\delta\le\alpha$.

Clearly, any nontrivial run of algorithm $A$ (i.e., one in which at least one
message is sent) requires at least one node to behave as an initiator in the
first of its calls to procedure $\textit{Fire}$. Henceforth, we let $m\le n$ be
the number of nodes that do this, that is, the number of initiators in the run.
It should also be clear, by steps 1 and 2 of algorithm $A$, that setting the
initial value of $v_j$ to some number in the interval $[v^0,v^\mathrm{t}]$
ensures that $v_j$ remains in this interval perpetually (this guarantees that
the value to which probability $p$ is set in step 3 is always legitimate).
Likewise, it follows from steps 4 and 5 that, if initialized to some number in
the interval $[0,1]$, weight $w_{ij}$ remains constrained to lie in this
interval for the whole run.

Any run of algorithm $A$ terminates eventually with probability $1$. That is,
there necessarily comes a time during the run at which no more messages are sent
and, from then on, no further processing occurs at the nodes. When graph $D$ is
strongly connected (i.e., a directed path exists from any of its nodes to any
other), then any firing during the run causes messages to be sent. Similarly,
any firing by a node that is not acting as initiator is preceded by
accumulating and/or depleting alterations to the node's potential, as messages
arrive, relative to the value it had initially or when the node last fired.
Message traffic, therefore, provides the essential backdrop against which to
conduct our study of information integration.

\section{Information integration}
\label{sec:infoint}

We consider $N$ discrete random variables, denoted by $X_1,X_2,\ldots,X_N$, each
taking values from the set $\{0,1\}$. All of our study on how information gets
integrated in a directed graph running the distributed algorithm $A$ of
Sec.~\ref{sec:algo} is based on attaching meaning to these variables, of which
there is one for each of $N\le n$ nodes of graph $D$, and to their
distributions. We do this later in this section and also in
Sec.~\ref{sec:results}. First, though, we establish the two indicators of
information integration that will be used.

For the sake of notational conciseness, we use $\mathbf{X}$ to denote the whole
sequence $X_1,X_2,\ldots,X_N$ of variables, and likewise
$\mathbf{x}\in\{0,1\}^N$ to denote one of the possible $2^N$ sequences of values
$x_1,x_2,\ldots,x_N\in\{0,1\}$, each for the corresponding variable.
Unambiguously, then, $\mathbf{X}=\mathbf{x}$ means that
$X_1=x_1,X_2=x_2,\ldots,X_N=x_N$. If $P(\mathbf{x})$ is the joint probability
that $\mathbf{X}=\mathbf{x}$, then we use $P_i(x_i)$ to denote the marginal
probability that $X_i=x_i$ for all $i\in\{1,2,\ldots,N\}$. Clearly, $P_i(x_i)$
is given by the sum of $P(\mathbf{x})$ over all $2^{N-1}$ possibilities for
$\mathbf{x}$ that leave the value of $X_i$ fixed at $x_i$.

Ultimately, our indicators of information integration are expressible in terms
of the Shannon entropy associated with the sequence $\mathbf{X}$ of variables
given the joint distribution $P$, or with each individual variable $X_i$ given
the corresponding marginal distribution $P_i$. This entropy gives, in
(information-theoretic) bits, a measure of how much unpredictability the
distribution embodies regarding the values of the variables. We denote the joint
entropy by $H(\mathbf{X})$ and each marginal entropy by $H_i(X_i)$. They are
given by the well-known formulae
\begin{equation}
H(\mathbf{X})=-\sum_{\mathbf{x}\in\{0,1\}^N}P(\mathbf{x})\log_2P(\mathbf{x})
\end{equation}
and
\begin{equation}
H_i(X_i)=-\sum_{x_i\in\{0,1\}}P_i(x_i)\log_2P_i(x_i).
\end{equation}
Recall that entropy is a function of the distribution and is maximized when the
distribution is uniform over its domain. Thus, $0\le H(\mathbf{X})\le N$ and
$0\le H_i(X_i)\le 1$.

The way our indicators become expressed as combinations of entropies is through
another fundamental information-theoretic notion, that of the relative entropy,
or Kullback-Leibler (KL) divergence, of two distributions \cite{mw}. Given two
joint distributions $P$ and $Q$ over the same set of $N$ variables as above,
the KL divergence of $P$ relative to $Q$, here denoted by $D(P,Q)$, is given by
\begin{equation}
D(P,Q)=\sum_{\mathbf{x}\in\{0,1\}^N}
P(\mathbf{x})
\log_2\frac{P(\mathbf{x})}{Q(\mathbf{x})},
\label{eq:kl}
\end{equation}
provided $Q(\mathbf{x})>0$ whenever $P(\mathbf{x})>0$. We have $D(P,Q)=0$ if and
only if $P$ and $Q$ are the same distribution. Otherwise $D(P,Q)>0$, so the KL
divergence functions as a measure of how different the two distributions are
[though, in general, $D(P,Q)\neq D(Q,P)$].

\subsection{Information gain}

The first of our two indicators, information gain, is the KL divergence of $P$
relative to $Q$ when the latter reflects a state of maximum unpredictability
regarding the values of the $N$ variables. That is, we use $Q(\mathbf{x})=1/2^N$
for all $\mathbf{x}\in\{0,1\}^N$. We denote information gain by
$G(\mathbf{X})$ and it follows from Eq.~(\ref{eq:kl}) that
\begin{equation}
G(\mathbf{X})=N-H(\mathbf{X}).
\label{eq:ig}
\end{equation}
Evidently, $0\le G(\mathbf{X})\le N$.

A marginal version of information gain for $X_i$ can also be defined by
recognizing that $Q_i(0)=Q_i(1)=0.5$. Denoting this marginal information gain by
$G_i(X_i)$, we have $G_i(X_i)=D(P_i,Q_i)$, whence
\begin{equation}
G_i(X_i)=1-H_i(X_i)
\label{eq:mig}
\end{equation}
and $0\le G_i(X_i)\le 1$.

Our use of information gain will be based on letting $N$ be the number of nodes
in the graph's GSCC, that is, one variable per node in the GSCC of graph $D$.
Moreover, $P_i(1)$ will be the probability that node $i$ receives at least one
message during a run of algorithm $A$ on $D$. Similarly, $P(\mathbf{x})$ will be
the probability that every node $i$ for which $x_i=1$ (and no other node)
receives at least one message during the run.

\subsection{Total correlation}

Our second indicator uses $Q(\mathbf{x})=\prod_{i=1}^NP_i(x_i)$ for all
$\mathbf{x}\in\{0,1\}^N$. That is, it addresses the question of how far
the variables $X_1,X_2,\ldots,X_N$ are from being independent from one another
relative to $P$. Given this choice for the joint distribution $Q$, the KL
divergence $D(P,Q)$ becomes what is known as the total correlation among the $N$
variables \cite{w60}, henceforth denoted by $C(\mathbf{X})$. It follows from
Eq.~(\ref{eq:kl}) that
\begin{equation}
C(\mathbf{X})=\sum_{i=1}^NH_i(X_i)-H(\mathbf{X})
\label{eq:tc}
\end{equation}
\footnote{The total correlation of $N=2$ variables coincides with their mutual
information \cite{lr09}. In general, though, the two quantities are markedly
different, since mutual information generalizes the $N=2$ case of
Eq.~(\ref{eq:tc}) differently \cite{h80}.}. Like entropy, total correlation is
expressed in bits and is a function of the joint distribution $P$. It is
maximized whenever $P$ assigns zero probability to all but two of the members of
$\{0,1\}^N$: if $\mathbf{x}$ and $\mathbf{y}$ are the two exceptions, then
maximization occurs if $\mathbf{x}$ and $\mathbf{y}$ are complementary value
assignments to the variables (that is, for all $i$ it holds that $x_i=0$ if and
only if $y_i=1$) and moreover $P(\mathbf{x})=P(\mathbf{y})=0.5$. Under these
conditions, clearly $H_i(X_i)=1$ for all $i$ and $H(\mathbf{X})=1$. Therefore,
$0\le C(\mathbf{X})\le N-1$.

By Eqs.~(\ref{eq:ig})--(\ref{eq:tc}), we have
$C(\mathbf{X})=G(\mathbf{X})-\sum_{i=1}^NG_i(X_i)$. That is, total correlation
is the amount of information gain that surpasses the total gain provided by the
variables separately. Equivalently, information gain comprises total correlation
and the total marginal gain $\sum_{i=1}^NG_i(X_i)$, i.e.,
\begin{equation}
G(\mathbf{X})=C(\mathbf{X})+\sum_{i=1}^NG_i(X_i).
\label{eq:gainparts}
\end{equation}

Our use of total correlation will also be based on letting $N$ be the number of
nodes in the GSCC of graph $D$. Moreover, both $P_i(1)$ and $P(\mathbf{x})$ will
have the same meanings as given above for information gain. We will also use the
ratio
\begin{equation}
r(\mathbf{X})=\frac{C(\mathbf{X})}{G(\mathbf{X})}
\label{eq:ratio}
\end{equation}
as an indicator of how conducive graph $D$ is, under algorithm $A$, to
generating information in the form of total correlation.

\subsection{Expected values}
\label{sec:infoint-exp}

Running the distributed algorithm $A$ of Sec.~\ref{sec:algo} on graph $D$ from a
set of initiators alters the edges' synaptic weights and, along with them, the
joint distribution $P$. As $P$ changes, so do $G(\mathbf{X})$ and
$C(\mathbf{X})$ and, in interpreting the results of Sec.~\ref{sec:results}, it
will be useful to have $G(\mathbf{X})$ and $C(\mathbf{X})$ values against which
to gauge the values that we obtain. Using the maximum values given above for
each quantity is of little meaning, since they occur only at finitely many
possibilities for $P$ while $P$ itself varies over a continuum of possibilities.

We then look at the expected value of either quantity as $P$ varies. To do so,
we first note that specifying $P$ is equivalent to specifying $2^N$ numbers in
the interval $[0,1]$, provided they add up to $1$. In other words, $P$ can be
identified with each and every point of the standard simplex in
$2^N$-dimensional real space. Calculating the expected value of either
$G(\mathbf{X})$ or $C(\mathbf{X})$ over this simplex requires the choice of a
density function and then an integration over the simplex. Given the complexity
of both the distributed algorithm and the structure of $D$, it seems unlikely
that a suitable density function can be derived. Moreover, even if we assume the
uniform density instead, there is still the task of integrating $G(\mathbf{X})$
and $C(\mathbf{X})$ over the simplex, which to the best of our knowledge can
be done analytically for $G(\mathbf{X})$, through the expected value of
$H(\mathbf{X})$ (cf.\ \cite{c95} and references therein), but not for
$C(\mathbf{X})$.

For sufficiently large $N$, it follows from the formula in \cite{c95} that the
expected value of $H(\mathbf{X})$ over the simplex using the uniform density
tends to $N-(1-\gamma)/\ln 2$, where $\gamma\approx 0.57722$ is the
Euler constant \cite{mw}. Therefore, by Eq.~(\ref{eq:ig}) the expected value of
$G(\mathbf{X})$ tends to the constant $(1-\gamma)/\ln 2\approx 0.6$. Similarly,
it follows from Eq.~(\ref{eq:tc}) that $0.6$ can also be taken as an approximate
upper bound on the expected value of $C(\mathbf{X})$. We also know from
\cite{c95} that, under these same conditions, $H(\mathbf{X})$ is tightly
clustered about the mean, and thus so is $G(\mathbf{X})$.

\section{Computational results}
\label{sec:results}

The methodology we follow in our computational experiments is entirely analogous
to the one introduced in \cite{nb10}. The central entity in this methodology is
a run of algorithm $A$ of Sec.~\ref{sec:algo}, using $v^0=-15$,
$v^\mathrm{t}=0$, $\delta=0.0002$, and $\alpha=0.04$ at all times. A run is
started by $m=50$ initiators chosen uniformly at random and progresses until
termination. These values of $\delta$ and $\alpha$ are the same that in
\cite{nb10} were shown to allow the synaptic weights to become distributed as
observed experimentally. As for $v^0$, $v^\mathrm{t}$, and $m$, their values
only regulate the traffic of causally disconnected messages in the graph and
therefore only influence how early in a sequence of runs global properties can
be expected to emerge.

For a fixed graph $D$, first we decide for each of the $N$ nodes in the graph's
GSCC whether it is to be excitatory or inhibitory. This is done uniformly at
random, provided no two inhibitory nodes are directly connected to each other.
We use the widely accepted proportion of $20\%$ for the number of inhibitory
nodes in $D$ \cite{abeles91,ananthanarayanan07}. All runs on graph $D$ operate
on this fixed set of inhibitory nodes. Then we choose initial node potentials
and synaptic weights uniformly at random from the intervals $[v^0,v^\mathrm{t}]$
and $[0,1]$, respectively. We group all runs on graph $D$ into sequences. The
first run in a sequence starts from the initial node potentials and synaptic
weights that were chosen for the graph. Each subsequent run starts from the node
potentials and synaptic weights left by the previous run. We use $50\,000$
sequences for each graph $D$, each sequence comprising $10\,000$ runs. We adopt
eleven observational checkpoints along the course of each sequence. The first
one occurs right at the beginning of the sequence, before any run takes place,
so node potentials and synaptic weights are still the ones chosen randomly. The
remaining ten checkpoints occur each after $1\,000$ additional runs in the
sequence.

The purpose of each checkpoint is to allow the joint distribution $P$ of the
variables $X_1,X_2,\ldots,X_N$ to be estimated and, based on it, the calculation
of information gain $G(\mathbf{X})$ and total correlation $C(\mathbf{X})$. Since
the marginal $P_i(1)$ is to reflect the probability that node $i$ receives at
least one message during a run, what is done at each checkpoint is to observe
the message propagation patterns that take place on graph $D$ as algorithm $A$
is executed on it. We do this by resorting to $100$ side runs of the algorithm,
each beginning with the choice of a new set of $m$ initiators uniformly at
random and starting from the node-potential and synaptic-weight values that are
current at the checkpoint. At the end of all side runs, the main sequence of
runs is resumed from these same values. For $c=1,2,\ldots,11$, the joint
distribution $P$ corresponding to the $c$th checkpoint can then be estimated
from the overall number of side runs, which is $5\times 10^6$. For the purpose
of averaging the resulting $G(\mathbf{X})$ and $C(\mathbf{X})$ values, multiple
instances of graph $D$ are needed. This is so in order to account for structural
variations and variations in the excitatory/inhibitory character of each node
(in case the graphs come from sampling from a random-graph model), and for
variations in the initial node potentials and synaptic weights (in all cases,
including the single case in which the structure of $D$ is deterministic;
cf.~below).

Estimating $P$ at the $c$th checkpoint proceeds as follows. After each side run
of $A$, the point $\mathbf{x}\in\{0,1\}^N$ such that $x_i=1$ if and only if node
$i$ received at least one message during the run has its number of occurrences
increased by $1$. Straightforward normalization yields $P(\mathbf{x})$ after all
sequences have reached that checkpoint on the graph in question and the
corresponding side runs have terminated. This poses a somewhat severe storage
problem, since for each $D$ we execute the sequences one after the other while
handling the graphs in parallel (on different processors). Therefore, the
accumulators corresponding to the various members of $\{0,1\}^N$ that are
actually observed have to be stored concomitantly for all eleven checkpoints.
There is no choice but to use external (i.e., disk-based) storage in this case,
which is heavily taxing with respect to how long it takes to complete
everything. So the number $5\times 10^6$ of side runs per checkpoint per graph
cannot in practice be made substantially larger. This number, after multiplied
by the number of graphs in use, is also an upper bound on how many members of
$\{0,1\}^N$ can be observed per checkpoint, so not being able to increase it
means that the number of nodes $n$ cannot be too large, either. All the results
we give henceforth are then for $n=100$.

We consider three different types of graph. They are referred to as
type-(i)--(iii) graphs, as follows:

(i) First is the random-graph model introduced in Sec.~\ref{sec:model} as the
structural component of our cortical model. Generating $D$ from this
random-graph model starts with placing the $n$ nodes on the surface of a
radius-$1$ sphere uniformly at random and then selecting, for each node, its
out-degree and its out-neighbors. The choice of $\lambda=-1$ explained in
Sec.~\ref{sec:model} is specific of the $n=100$ case and yields $N\approx 90$.
Also, for this value of $n$ the expected in- or out-degree in $D$ is about
$3.7$. Out-degrees are by construction distributed as a power law, whereas the
distribution of in-degrees has been found to be similar to the Poisson
distribution (i.e., concentrated near the mean) \cite{nb10}.

(ii) Another random-graph model that we use is the generalization of the
Erd\H{o}s-R\'{e}nyi model to the directed case \cite{k90}. Given the desired
expected in- or out-degree, denoted by $z$, generating $D$ places a directed
edge from node $i$ to node $j\neq i$ with probability $z/(n-1)$. The resulting
in- and out-degree distributions approach the Poisson distribution of mean $z$.
If $z>1$, the graph's GSCC encompasses nearly all the graph with high
probability; that is, $N\approx 100$. For consistency with our cortical model,
we use $z=3.7$.

(iii) At the other extreme from our cortical model are the graphs whose
structure is deterministic. We use what seems to be the simplest possible
structure that ensures a strongly connected $D$ with a fixed in- or out-degree
equal to $\lceil 3.7\rceil=4$ for every node, the directed circulant graph
\cite{lpw01} generated by the integers in the interval $[1,4]$. If we assume
that the nodes are numbered $0$ through $n-1$, then node $i$ has four
out-neighbors, nodes $i+1$ through $i+4$, where addition is modulo $n$. For
$n=100$, the $20$ inhibitory nodes are necessarily equally spaced around the
directed cycle that traverses the nodes in the order $0,1,\ldots,n-1,0$, lest
there be a connection between two inhibitory nodes. We have $N=100$.

\begin{figure*}
\centering
\scalebox{0.50}{\includegraphics{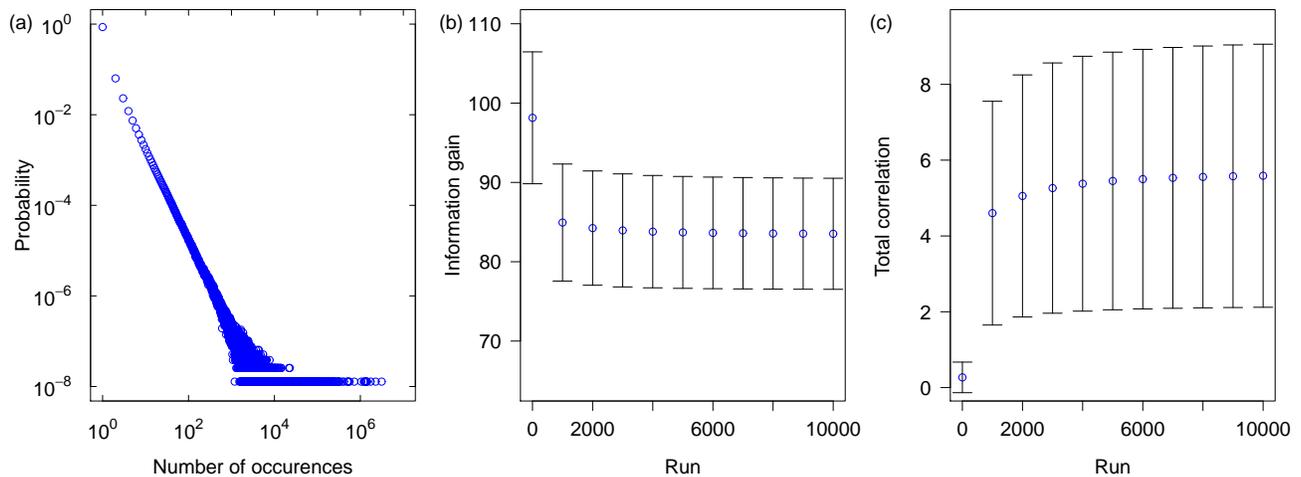}}
\caption{(Color online) Results for type-(i) graphs: (a) the probability that
a randomly chosen member of $\{0,1\}^N$ appearing in the side runs of the last
checkpoint for some graph occurs a certain number of times; (b) the average
value of $G(\mathbf{X})$ at each of the checkpoints; (c) the average value of
$C(\mathbf{X})$ at each of the checkpoints.}
\label{fig:figure2}
\end{figure*}

\begin{figure*}
\centering
\scalebox{0.50}{\includegraphics{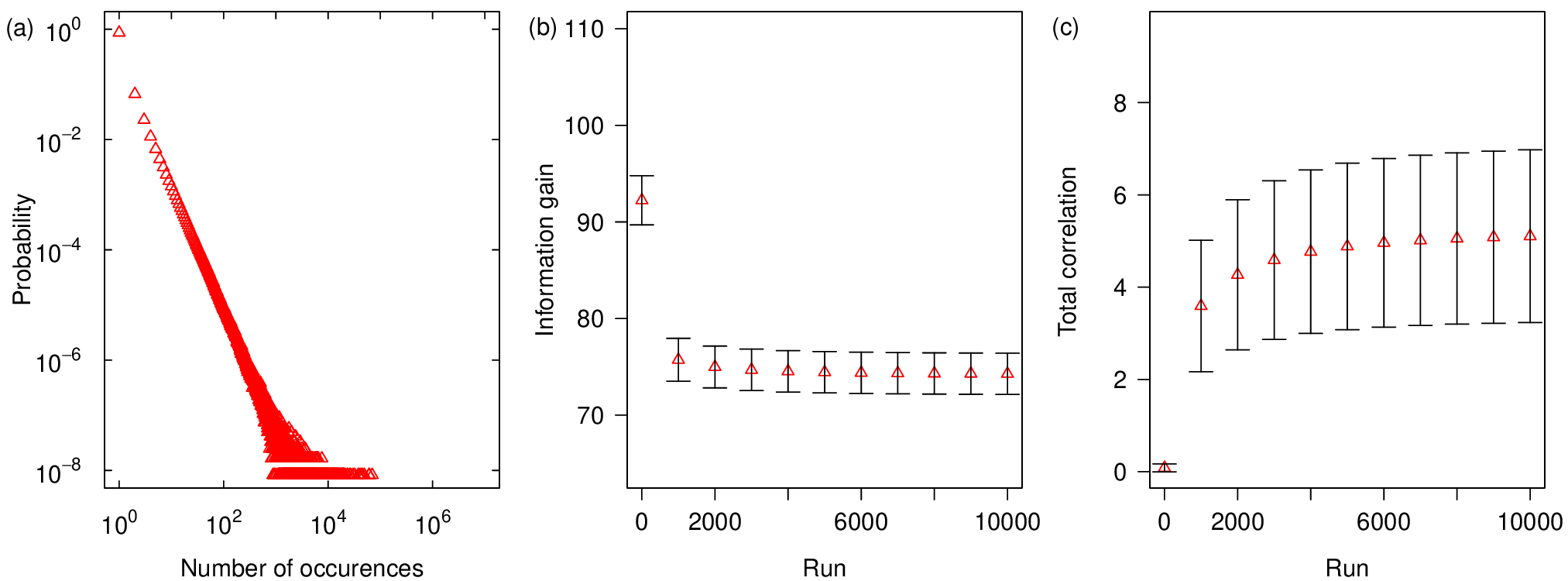}}
\caption{(Color online) Results for type-(ii) graphs: (a) the probability that
a randomly chosen member of $\{0,1\}^N$ appearing in the side runs of the last
checkpoint for some graph occurs a certain number of times; (b) the average
value of $G(\mathbf{X})$ at each of the checkpoints; (c) the average value of
$C(\mathbf{X})$ at each of the checkpoints.}
\label{fig:figure3}
\end{figure*}

\begin{figure*}
\centering
\scalebox{0.50}{\includegraphics{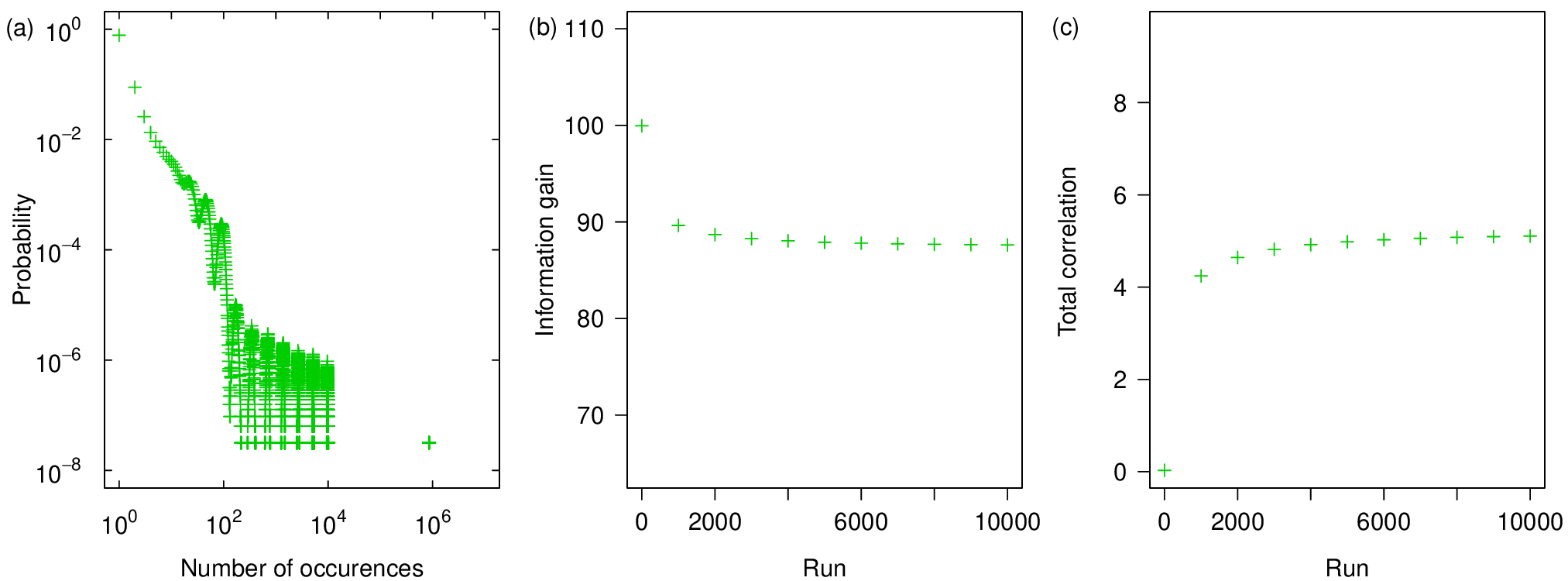}}
\caption{(Color online) Results for type-(iii) graphs: (a) the probability that
a randomly chosen member of $\{0,1\}^N$ appearing in the side runs of the last
checkpoint for some graph occurs a certain number of times; (b) the average
value of $G(\mathbf{X})$ at each of the checkpoints; (c) the average value of
$C(\mathbf{X})$ at each of the checkpoints.}
\label{fig:figure4}
\end{figure*}

All our results are given for $50$ graphs of each of types (i)--(iii) and appear
in Figs.~\ref{fig:figure2}--\ref{fig:figure4}, respectively. The (a) panels in
these figures give the probability distributions for the number of occurrences
of those members of $\{0,1\}^N$ that do appear in at least one side run on at
least one of the $50$ graphs for each graph type at the eleventh checkpoint.
After averaging over the appropriate $50$ graphs for each type, these members
number $1\,733$ for type-(i) graphs, $4\,756$ for type-(ii) graphs, and $1\,033$
for type-(iii) graphs. These illustrate the point, raised above, that the need
to limit the total number of side runs per graph does indeed have an impact on
how capable our methodology is to probe inside the set of all $2^N$ value
assignments to the $N$ variables. In fact, the absolute majority of assignments
are never encountered. As for the others, the probability of encountering them
an increasing number of times decays as a power law [less so for type-(iii)
graphs].

The (b) and (c) panels in the three figures are used to show the average
information gain $G(\mathbf{X})$ and total correlation $C(\mathbf{X})$,
respectively, over the $50$ graphs of the corresponding graph type at each of
the eleven checkpoints. Error bars are omitted from the (b) and (c) panels of
Fig.~\ref{fig:figure4} because the corresponding standard deviations are
negligible. Because neither $G(\mathbf{X})$ nor $C(\mathbf{X})$ can surpass the
number $N$ of nodes in the graph's GSCC, and considering that not all graphs
across the three graph types have the same GSCC size, all the data plotted in
the (b) and (c) panels of Figs.~\ref{fig:figure2}--\ref{fig:figure4} are
normalized to this size. The latter, in turn, can be taken as $0.9n$ for
type-(i) graphs and $n$ for type-(ii) and type-(iii) graphs. Our normalization
procedure, therefore, has been to divide all $G(\mathbf{X})$ and $C(\mathbf{X})$
values for type-(i) graphs by $0.9$, leaving them unchanged for graphs of the
other two types.

\begin{figure}
\centering
\scalebox{0.50}{\includegraphics{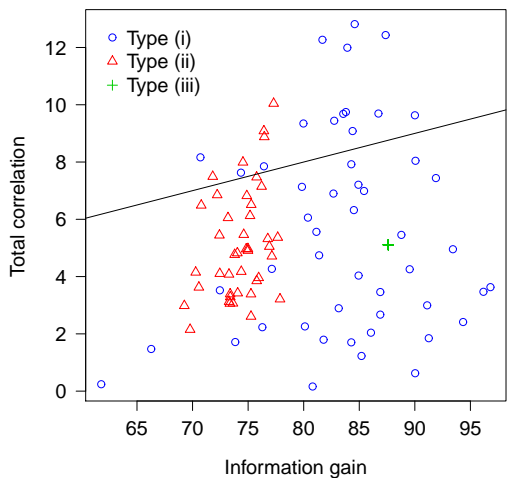}}
\caption{(Color online) A scatter plot of the $150$ different graphs used, $50$
for each of types (i)--(iii). Each graph is represented by its information gain
$G(\mathbf{X})$ and its total correlation $C(\mathbf{X})$ at the last
checkpoint. The straight line below which most type-(ii) and all type-(iii)
graphs are found goes through the origin with slope $0.1$.}
\label{fig:figure5}
\end{figure}

A different perspective on the results shown in
Figs.~\ref{fig:figure2}--\ref{fig:figure4} is given in Fig.~\ref{fig:figure5},
which presents a scatter plot of all $150$ graphs of the three types, each
represented by its information gain and its total correlation at the last
checkpoint. In this figure, the same normalization described above has also been
used.

\section{Discussion}
\label{sec:disc}

The (b) panels in Figs.~\ref{fig:figure2}--\ref{fig:figure4} demonstrate that
the $G(\mathbf{X})$ averages, after a sharp decrease from the first checkpoint
to the second, keep on decreasing steadily along the runs until stability is
eventually reached. A similar trend is seen in the (c) panels with regard to the
$C(\mathbf{X})$ averages, now with increases. (We note that reaching stability
in either case is on a par with what, in \cite{nb10}, we showed to happen with
the distribution of synaptic weights under the same cortical model. That is,
despite the continual modification of the weights as the dynamics goes on, their
distribution reaches a steady state.) With the exception of
Fig.~\ref{fig:figure4}, standard deviations can be significant all along the
runs, particularly with regard to total correlation.

Given this variability, the plot in Fig.~\ref{fig:figure5} is important and also
quite revealing. First of all, it helps corroborate what the (b) and (c) panels
of Figs.~\ref{fig:figure2}--\ref{fig:figure4} already say about algorithm $A$,
which essentially is what drives the system toward the eventual joint
distribution $P$ over the variables $X_1,X_2,\ldots,X_N$ that is used to compute
$G(\mathbf{X})$ and $C(\mathbf{X})$ for each graph. As we discussed in
Sec.~\ref{sec:infoint-exp}, should all possibilities for $P$ be equally likely,
$G(\mathbf{X})$ would have a mean value of about $0.6$ over all these
possibilities and would moreover be tightly clustered about this mean. The
$G(\mathbf{X})$ values appearing in Fig.~\ref{fig:figure5} demonstrate that
algorithm $A$, independently of which graph type is used, completely subverts
the uniformity hypothesis for $P$ and leads the system to generate information
in amounts that surpass the $0.6$ mark very significantly. This holds also with
regard to the $C(\mathbf{X})$ values in Fig.~\ref{fig:figure5}, whose mean under
uniformly weighted $P$'s is also bounded from above by $0.6$.

Figure~\ref{fig:figure5} also allows us to investigate, for each graph at the
last checkpoint, how its information gain $G(\mathbf{X})$ and total correlation
$C(\mathbf{X})$ are related to each other. The simplest case is that of
type-(iii) graphs, whose topology and inhibitory-node positions are fixed at all
times. In this case, the stochasticity of initial node potentials and synaptic
weights, as well as of the functioning of algorithm $A$, are insufficient to
yield any significant variation in $G(\mathbf{X})$ or $C(\mathbf{X})$ values.
Next are the type-(ii) graphs, for which neither topology nor the placement of
inhibitory nodes is the same for all graphs. What we see as a result is
significantly more variation in $G(\mathbf{X})$ and $C(\mathbf{X})$ values, but
with very few exceptions all $50$ graphs are still discernibly clustered
relative to one another. Type-(i) graphs, finally, with their dependence of
topology upon both a power-law-distributed out-degree and the random placement
of nodes on a sphere, display $G(\mathbf{X})$ and $C(\mathbf{X})$ values that
are spread over a significantly larger domain.

Aside from such broad qualitative statements, it seems hard to discriminate
among the three graph types by examining either $G(\mathbf{X})$ or
$C(\mathbf{X})$ values alone, even though there are type-(i) graphs for which
$G(\mathbf{X})$ is greater than for any graph of the other two types, the same
holding for $C(\mathbf{X})$. One simple, though effective, alternative is to
resort to the ratio $r(\mathbf{X})$ defined in Eq.~(\ref{eq:ratio}). This ratio
gives the fraction of all the information generated by the system that
corresponds to total correlation, that is, the fraction that corresponds to
information that depends on integration among the variables. Once we adopt this
metric, then the meaning of Fig.~\ref{fig:figure5} becomes clearer: although all
three types of graph are capable of providing significant information gain and
total correlation, only type-(i) graphs (those at the basis of our cortical
model) seem capable of providing an abundance of instances for which
$r(\mathbf{X})$ is higher than for most type-(ii) graphs and all type-(iii)
graphs.

By its very definition, the ratio $r(\mathbf{X})$ for a given graph can be
regarded as an indicator of how efficient that graph is, under algorithm $A$, at
integrating information. Graphs for which $r(\mathbf{X})$ is higher than for
others do a better job in the sense that, of all the information that they
generate, a higher fraction corresponds to information that emerges out of the
integration among their constituents. What our results indicate is that type-(i)
graphs, based as they are on a random-graph model, can be instantiated to
specific graphs that are often more efficient than those of the other two types.
The straight line drawn across Fig.~\ref{fig:figure5} has slope $0.1$ and can be
used as an example discriminator on the $150$ graphs represented in the figure
with respect to efficiency. Specifically, all graphs above it are such that
$r(\mathbf{X})>0.1$. The overwhelming majority of them are type-(i) graphs.

Justifying this behavior in terms of graph structure is still something of an
open problem, though. We believe the justification has to do with the existence
of hubs in type-(i) graphs, since one of their effects is to shorten distances,
but whatever it is remains to be made precise. One might also think that the way
in- and out-degrees get mixed in type-(i) graphs could also constitute a line of
explanation, especially because these graphs have markedly different in- and
out-degree distributions. If this were the case then it might be reflected in
the statistical properties of these graphs' assortativity coefficient
\cite{n03}, which is the Pearson correlation coefficient of the edges' remaining
out-degrees on the tail sides and remaining in-degrees on the head sides.
Recall, however, that generating type-(i) graph instances, just like generating
type-(ii) instances, makes no reference whatsoever to node degrees when deciding
which nodes are to be joined by a given edge, so the expected assortativity
coefficient of graphs of either type is zero in the limit of a formally infinite
number of nodes \footnote{This holds also for the more recent alternative
definitions of the assortativity coefficient given in \cite{ppz11}. Contrasting
with the definition in \cite{n03}, these correlate in-degrees with in-degrees or
out-degrees with out-degrees.}, quite unlike the fixed structure of a type-(iii)
graph (for which the assortativity coefficient is $1$). We have verified that
this holds by resorting to $1\,000$ independent instances of type-(i) graphs for
$n=100$ and keeping the calculations inside each graph's GSCC. In this
experiment we also found that the standard deviation of the assortativity
coefficient is of the order of $10^{-2}$, so there is little variation from
graph to graph.

\begin{figure}
\centering
\scalebox{0.50}{\includegraphics{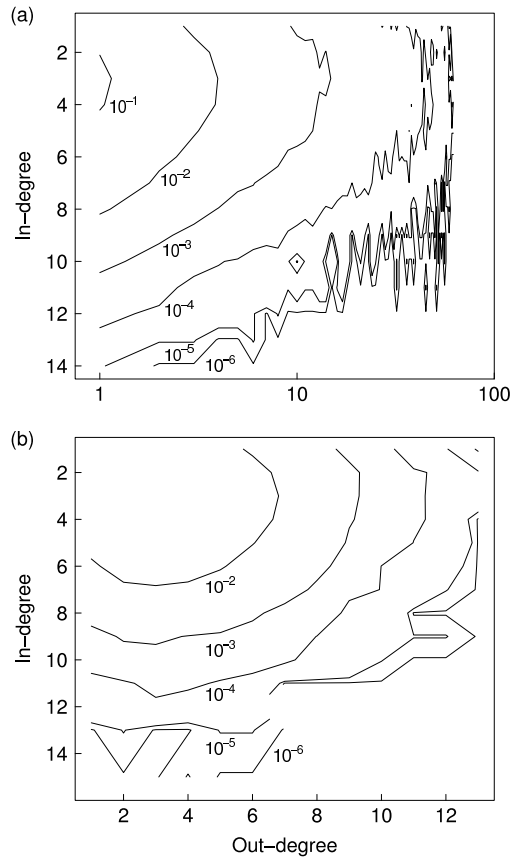}}
\caption{Contour plots of the joint distribution of a node's in- and out-degree
for type-(i) (a) and type-(ii) (b) graphs. Data are averages over $1\,000$
graphs of each type for $n=100$, always restricted to each graph's GSCC.}
\label{fig:figure6}
\end{figure}

A possible route to analyzing the role of hubs in giving rise to efficient
information integration predominantly in type-(i) graphs may be to study the
joint distributions of in- and out-degrees. These distributions are shown in
Fig.~\ref{fig:figure6}, in the form of contour plots, for type-(i) and type-(ii)
graphs with $n=100$. In the figure, all data are averages over the inside of
each graph's GSCC, so in- and out-degrees are expected to be no larger than
$90$. While for type-(ii) graphs, in reference to part (b) of the figure, we
expect no nodes to exist whose in- and out-degrees differ from each other
significantly, the case of type-(i) graphs is substantially different. First of
all, the data in part (a) of the figure reveal that the most common combination
of in- and out-degree at a node is that in which the node has a small number of
in-neighbors (between $2$ and $4$) and an even smaller number (in fact, no more
than $2$) of out-neighbors. Such nodes function somewhat as type of
concentrator, meaning that whenever they fire in the wake of the accumulation of
signaling from its in-neighbors the resulting signal affects at most two other
nodes. Hubs occur in the opposing end of this spectrum. If we take a node to be
a hub when it has, say, at least $50\%$ of the other nodes as out-neighbors,
then we see that, though hubs are very rare, when they occur they function as a
type of disseminator: when they fire, they affect substantially more nodes than
the handful of in-neighbors whose accumulated signals led to the firing itself.
Perhaps it is the combination of these two types of behavior, viz.\ an abundance
of concentrators and the occasional occurrence of a disseminator, that explains
the information-integration behavior we have observed for type-(i) graphs
\footnote{Similar arrangements, in the totally distinct context of evolutionary
graph dynamics, can be shown to result in strong amplification of the
probability that new mutations become widely spread \cite{lhn05}. It remains to
be seen whether the analogy between the two contexts goes any further than this
similarity.}. We expect that more research will clarify whether this is the
case.

\section{Concluding remarks}
\label{sec:concl}

We have introduced a network-algorithmic framework to study the emergence of
information integration in directed graphs. Our original inspiration was the IIT
of \cite{balduzzi08}, but the resulting framework departs significantly from IIT
in several aspects, most notably the adoption of the asynchronous distributed
algorithm of \cite{nb10} to simulate neuronal processing and signaling, and the
use of binary variables, one for each node in the graph's GSCC, to signify
whether nodes are reached by at least one message during a run of the algorithm.
These variables have been the basis on which two information-theoretic
quantities can be computed, namely information gain $G(\mathbf{X})$ and total
correlation $C(\mathbf{X})$. Given the graph's structure, the former of these
indicates how much information the system generates, under the distributed
algorithm, from an initial state of total uncertainty. The latter, in turn,
indicates how much of this information is integrated, as opposed to the
information that each node generates locally, independently of all others,
denoted by $G_i(X_i)$ for node $i$. For $N$ the number of variables, these
quantities are related by Eq.~(\ref{eq:gainparts}). This equation, with
hindsight, can nowadays be seen to have been present, at least qualitatively, in
most information-theoretic views of system organization and structure
(cf.~\cite{r52} for an early example). If we stick with the basic premise of
IIT, that consciousness and some form of information integration are to be
equated, then what the equation says is that, of all the information that the
system generates [$G(\mathbf{X})$], some reflects conscious processing
[$C(\mathbf{X})$] and some unconscious processing [$\sum_{i=1}^NG_i(X_i)$].

We have studied the behavior of information gain and total correlation for a
variety of graphs. These have included (i) the random graphs that, as we know
from our earlier study in \cite{nb10}, reproduces some experimentally observed
cortical properties; (ii) random graphs with Poisson-distributed in- and
out-degrees; and (iii) the deterministically structured circulant graphs. While
we have found that many instances of these graphs are capable of generating
comparable amounts of both information gain and total correlation, those that do
so efficiently (i.e., with a comparatively high $C(\mathbf{X})/G(\mathbf{X})$
ratio) are very predominantly of type (i). In the context of regarding
information integration as consciousness, this seems to provide further evidence
that the cortical model introduced in \cite{nb10} can indeed be useful as a
framework for the study of cortical dynamics. Another interesting aspect, now
related to the actual $G(\mathbf{X})$ and $C(\mathbf{X})$ values we have
observed, is that the latter are much lower than the former. Once again, though,
the association of consciousness with integrated information seems illuminating,
since the overwhelming majority of all processing in the brain is believed to
occur unconsciously \cite{d99}.

As it stands, our framework is only capable of handling relatively small graphs.
The main difficulty is that we need to organize statistics of the frequency of
occurrence of the various members of $\{0,1\}^N$ that appear in the runs as they
elapse and this requires huge amounts of input/output operations on external
storage. With current technology, the results presented in
Sec.~\ref{sec:results} can require up to three weeks to complete for each graph.
And while the potential for parallelism is very great, normally one is also
limited on the number of processors one can count on. An important direction in
which to continue this research is to address these computational limitations.
Success here will immediately facilitate important further research on crucial
aspects of our conclusions, for example those related to how our finds scale
with increasing system size. We also find our underlying cortical model, with
its structural and algorithmic components, to be suitable for the undertaking of
investigations on entirely different fronts. One possibility of interest to us
is how to look for, and characterize, the emergence of certain oscillatory
patterns of cortical activity \cite{cgkckwc00}. It seems that the central issue
is how to reconcile such oscillations with the inherently asynchronous character
of our model. This, however, remains open to further research.

\begin{acknowledgments}
We acknowledge partial support from CNPq, CAPES, and a FAPERJ BBP grant.
\end{acknowledgments}

\bibliography{infoint}
\bibliographystyle{apsrev}

\end{document}